\documentclass[11pt,twoside]{article}

\usepackage{asp2006}
\usepackage{epsf}
\usepackage{psfig}
\usepackage{lscape}
\usepackage{graphicx}

\newcommand{\vc}[1]{\textbf{\em #1}}

\newcommand{\pder}[2]{\frac{\partial #1}{\partial #2}}

\begin{document}
\title{Model of Reconnection of Weakly Stochastic Magnetic Field and its Testing}   
\author{A. Lazarian\altaffilmark{1}, E. Vishniac\altaffilmark{2}, and G. Kowal\altaffilmark{1,3}}   
\affil{$^{1}$Department of Astronomy, University of Wisconsin-Madison, USA (lazarian@astro.wisc.edu),\\
$^{2}$Department of Physics \& Astronomy, McMaster University, Canada,\\
$^{3}$Astronomical Observatory, Jagiellonian University, Krak\'ow, Poland}    

\begin{abstract} 
Astrophysical fluids are generically turbulent, which means that frozen-in magnetic fields are, at least, weakly stochastic. Therefore realistic studies of astrophysical magnetic reconnection should include the effects of stochastic magnetic field. In the paper we discuss and test numerically the Lazarian \& Vishniac (1999) model of magnetic field reconnection of weakly stochastic fields. The turbulence in the model is assumed to be subAlfvenic, with the magnetic field only slightly perturbed. The model predicts that the degree of magnetic field stochasticity controls the reconnection rate and that the reconnection can be fast independently on the presence or absence of anomalous plasma effects. For testing of the model we use 3D MHD simulations. To measure the reconnection rate we employ both the inflow of magnetic flux and a more sophisticated measure that we introduce in the paper. Both measures of reconnection provide consistent results. Our testing successfully reproduces the dependences predicted by the model, including the variations of the reconnection speed with the variations of  the injection scale of turbulence driving as well as the intensity of driving. We conclude that, while anomalous and Hall-MHD effects in particular circumstances may be important for the initiation of reconnection, the generic astrophysical reconnection is fast due to turbulence, irrespectively of the microphysical plasma effects involved. This conclusion justifies  numerical modeling of many astrophysical environments, e.g. interstellar medium, for which plasma-effect-based collisionless reconnection is not applicable.
\end{abstract}


\section{Reconnection of Weakly Stochastic Magnetic Field}   

Magnetic reconnection is a fundamental problem of astrophysical MHD (see e.g.,
Priest \& Forbes 2000 and references therein). Indeed, while the condition of magnetic field 
being frozen is well satisfied in the bulk of astrophysical plasma, one should wonder what
happens when magnetic flux tubes try to push through each other. This is a very important question. For instance, the commonly employed {\it dynamo} theory
allows a large-scale magnetic field to grow exponentially at the expense of small-scale turbulent energy. However, it has been known for some time that the back reaction of the field can suppress magnetic diffusion and defeat the dynamo. Indeed, if a parcel of fluid is threaded by a field loop in
equipartition with the turbulence, the loop acts like a rubber band returning the parcel to its origin. As a result, small-scale magnetic fields accumulate most of the energy, while the mean magnetic field saturates far below equipartition with the surrounding fluid. To enable free motions of the fluid parcels, as required
by the mean field dynamo, there should be a mechanism for cutting the
``rubber bands''. This mechanism is magnetic reconnection. Unfortunately, if reconnection happens at the rate allowed by generally accepted Sweet-Parker model (Parker 1957, Sweet 1958), it is far too slow. Turbulence would cause many magnetic reversals per parsec within the interstellar medium. Observations, on the contrary, show that magnetic field is coherent over the scales of hundreds of parsecs. This fact, as well as direct observations of Solar flares (which are generally accepted to be fed by reconnection), suggest that the rate of reconnection is many orders of magnitude more rapid than the Sweet-Parker theory suggests.

In general, if the reconnection were slow,
the change of magnetic topology did not happen and the interacting flux tubes would create knots, redistributing 
magnetic energy to small scales. This is the expected outcome for the Sweet-Parker reconnection, which is extremely slow and inadequate for nearly all astrophysical situations. 

Petscheck (1964) reconnection is an attempt to remedy the problem assuming that magnetic flux tubes get into contact over areas determined by microphysics and the reconnection regions open up. So far, the feasibility of Petscheck reconnection has been demonstrated in very restricted circumstances (see Shay \& Drake 1998), which, for instance, exclude interstellar medium and other important astrophysical environments like stars and many types of accretion disks. One of the sources of these restrictions arises from the requirement that the medium should be collisionless in a very special sense,
namely, that the extension of the current sheet should not exceed several dozens of electron mean free paths (see Yamada et al. 2006). 

\begin{figure}[!t]
\includegraphics[width=0.7\columnwidth]{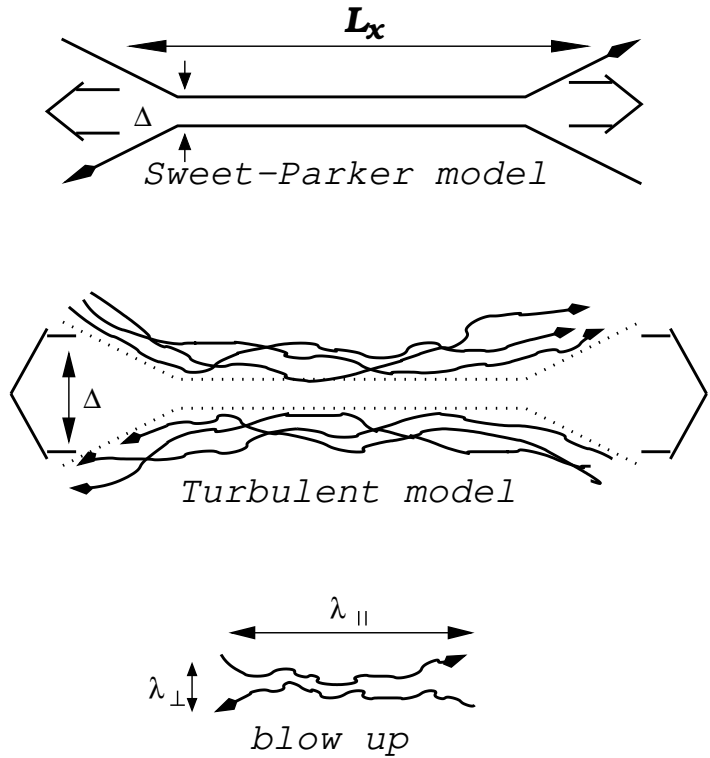}
\caption{{\it Upper plot}: 
Sweet-Parker model of reconnection. The outflow
is limited by a thin slot $\Delta$, which is determined by Ohmic 
diffusivity. The other scale is an astrophysical scale $L\gg \Delta$.
{\it Middle plot}: Reconnection of weakly stochastic magnetic field according to 
LV99. The model that accounts for the stochasticity
of magnetic field lines. The outflow is limited by the diffusion of
magnetic field lines, which depends on field line stochasticity.
{\it Low plot}: An individual small scale reconnection region. The
reconnection over small patches of magnetic field determines the local
reconnection rate. The global reconnection rate is substantially larger
as many independent patches come together.}
\label{recon}
\end{figure}

 Lazarian \& Vishniac (1999, henceforth LV99) proposed a model that naturally generalizes Sweet-Parker reconnection scheme for weakly turbulent magnetic fields. LV99 consider the case in which there exists a large scale, well-ordered magnetic field, of the kind that is normally used as
a starting point for discussions of reconnection\footnote{The LV99 model has nothing to do with the concept of turbulent magnetic diffusivity, which by some authors to justify the mean field dynamo. The latter concept requires bending of magnetic fields on very small scales, which is prohibited on energetic grounds for any dynamically important field.} (see Fig.~\ref{recon}). The difference with the Sweet-Parker reconnection arises both  from the fact that in 3D generic configuration (note that 3D is essential and the effect of fast reconnection is absent in 2D) many independent patches of magnetic field get into contact and undergo reconnection. In addition, the outflow of plasma and shared magnetic flux happens not over a microscopically narrow region determined by Ohmic diffusion, but through a substantially wider region determined by field wandering. LV99 showed that magnetic reconnection gets fast when these two effects are taken into account. In fact, the LV99 reconnection rate is
\begin{equation}
V_{rec} = V_A \min \left[ \left( \frac{L}{l} \right)^{1/2}, \left( \frac{l}{L} \right)^{1/2} \right] \left( \frac{V_l}{V_A} \right)^{1/2},
\label{eq:constraint}
\end{equation}
where $V_A$ is the Alfv\'en speed, $L$ is the size of the system, $l$ is the
injection scale, and $V_l$ is the velocity amplitude at the injection scale. In
this relation, the reconnection speed is determined by the characteristics of
turbulence, namely, its strength and injection scale. Most importantly, this rate is determined
by magnetic field wandering and contains no explicit dependence on the Ohmic or anomalous resistivity.
For isotropic injection of energy when the injection velocity is less than $V_A$ the injection power $P_{inj}$ is proportional to $V_l^4$ (see LV99), which means that Eq.~(ref{eq:constraint}) predicts
$V_{rec}\sim P_{inj}^{1/2} l^{1/2}$, assuming that  $l<L$. This is the dependence that we are testing in the paper. 

Within this short paper we do not have space to discuss in detail neither LV99 model nor other approaches to reconnection attempted through years by various reseachers. More on these issues can be found in the paper by \citet{lazarian08}, where also the implications
of the LV99 model of reconnection are also discussed. We instead concentrate here on numerical
testing of the model. The initial numerical testing of LV99 model is presented in \cite{kowal08}, where 
the reconnection rate was measured as the inflow rate of unreconnected flux. Here we present a more sophisticated and more theory-justified approach to measuring of the reconnection rate and compare the results of the two approaches.
\section{Numerical Modeling of LV99 Reconnection}
\label{sec:model}

We use a higher-order shock-capturing Godunov-type scheme based on the
essentially non oscillatory (ENO) spacial reconstruction and Runge-Kutta (RK)
time integration \citep[see][e.g.]{delzanna03} to solve isothermal non-ideal MHD
equations,
\begin{eqnarray}
 \pder{\rho}{t} + \nabla \cdot \left( \rho \vc{v} \right) & = & 0, \label{eq:mass} \\
 \pder{\rho \vc{v}}{t} + \nabla \cdot \left[ \rho \vc{v} \vc{v} + p_T I - \frac{\vc{B} \vc{B}}{4 \pi} \right] & = & \vc{f}, \label{eq:momentum} \\
 \pder{\vc{A}}{t} + \vc{E} & = & 0 \label{eq:induction},
\end{eqnarray}
where $\rho$ and $\vc{v}$ are plasma density and velocity, respectively,
$\vc{A}$ is vector potential, $\vc{E} = - \vc{v} \times \vc{B} + \eta \, \vc{j}$
is electric field, $\vc{B} \equiv \nabla \times \vc{A}$ is magnetic field,
$\vc{j} = \nabla \times \vc{B}$ is current density, $p_T = a^2 \rho + B^2 / 8
\pi$ is the total pressure, $a$ is the isothermal speed of sound, $\eta$ is
resistivity coefficient, and $\vc{f}$ represents the forcing term. We
incorporated the field interpolated constrained transport (CT) scheme
\citep[see][]{toth00} in to the integration of the induction equation to
maintain the $\nabla \cdot \vc{B} = 0$ constraint numerically.

Our initial magnetic field is a Harris current sheet of the form $B_x = B_{x0}
\tanh (y/\theta)$ with $B_{x0}= 1$. We use a uniform shear component $B_z =
B_{z0} = \mathrm{const}$ which varies between 0.0 and 1.0 depending in the
model. The initial setup is completed by setting the density profile from the
condition of the uniform total pressure $p_T(t=0) = \mathrm{const}$ and setting
the initial velocity to zero everywhere. The speed of sound is set to 4. In
order to study the resistivity dependence on the reconnection we vary the
resistivity coefficient $\eta$ between values $0.5\cdot10^{-4}$ and
$2\cdot10^{-3}$ which are expressed in dimensionless units. This means that the
velocity is expressed in units of Alfv\'en speed and time in units of Alfv\'en
time $t_A = L / V_A$, where $L$ is the size of the box. We initiate the magnetic
reconnection using a small perturbation of vector potential $\delta A_z = B_{x0}
\cos(2 \pi x) \exp[-(y/d)^2]$ to the initial configuration of $A_z(t=0)$. The
parameter $d$ describes the thickness of the perturbed region.

Numerical model of the LV99 reconnection is evolved in a box with open boundary
conditions which we describe in the next sub-section. The box has sizes $L_x =
L_z = 1$ and $L_y=2$ with the resolution 256x512x256. It is extended in
Y-direction in order to move the inflow boundaries far from the injection
region. This minimizes the influence of the injected turbulence on the inflow.

In our model we drive turbulence using a method described by \citet{alvelius99}.
It is implemented in spectral space. The input energy is concentrated with a
Gaussian profile around a wave vector corresponding to the injection scale
$l_{inj}$. The randomness in the time makes the force neutral in the sense that
it does not directly correlate with any of the time scales of the turbulent flow
and it also makes the power input determined solely by the force-force
correlation. The driving is completely solenoidal, which means that it does not
produce density fluctuations.

\section{Reconnection Rate Measures}
\label{sec:rec_rate}

One can measure the reconnection rate by averaging the inflow velocity $V_{in}$
divided by the Alfv\'en speed $V_A$ over the inflow boundaries. In this way our
definition of the reconnection rate is
\begin{equation}
V_{rec} = \langle V_{in} / V_A \rangle_{S} = \int_{y=y_{min},y_{max}}{ \frac{\vec{V}}{V_A}  \cdot d\vec{S}},
\label{oldmeasure}
\end{equation}
where $S$ defines the XZ planes of the inflow boundaries. This measure is the most natural
definition of the reconnection rate in the case of the laminar reconnection (Petscheck or Sweet-Parker ones). In the presence of turbulence, however, we
could face an uncertainty arising from the fact that the turbulence driven in the center of the box, can
remove inflow magnetic flux before it reaches the diffusion region and undergo
the reconnection process. Whether or not this effect is important we may test only by using a different measure that takes the loss of unreconnected flux into account. 

Our basic approach is to start by considering a conserved quantity, the magnetic
flux $\Phi$. First, we consider the flux contained within a plane inside the
simulation volume. If $\hat{x}$ is the direction of the reconnecting field, then
we start by considering the time change of the net flux of $B_x$. It is
\begin{equation}
\partial_t \Phi = \oint \vc{E} \cdot d \vc{l} = \oint \left( \vc{v} \times \vc{B} - \eta \vc{j} \right) \cdot d \vc{l}
\end{equation}

If we evaluate the difference between the two sides we will find that it is not
zero. Here, the resistivity $\eta$ has to be included. 
Now we split the area of integration into two pieces, $A_+$ and $A_-$, defined
by the sign of $B_x$. The fluxes have different sign, thus to consider the reconnection, which decreases the absolute value of magnetic flux, we have to subtract them, i.e.
\begin{equation}
\partial_t \Phi_{+} - \partial_t \Phi_{-} = \partial_t \int |B_x| dA,
\end{equation}
which we can write explicitly in terms of line integrals around $A_+$ and $A_-$
\begin{equation}
\partial_t \left[ \int |B_x| dA \right] = \oint{\vc{E} \cdot d \vc{l}_{+}} - \oint{\vc{E} \cdot d \vc{l}_{-}} = \oint \mathrm{sign} (B_x) \vc{E} \cdot d \vc{l} + \int 2 \vc{E} \cdot d \vc{l}_\mathrm{interface},
\end{equation}
where $\vc{l}_\mathrm{interface}$ is the line separating $A_+$ and $A_-$. The
last term describes the mutual annihilation of positive and negative $B_x$ along
the line separating them. By definition, this is the reconnection rate. Note, that
this includes the motion of already reconnected flux lines through the plane of
integration. Rather than try to calculate it numerically we define the interface
term as $-2 V_{rec} |B_{x,\infty}| L_z$, where $|B_{x,\infty}|$ is the
asymptotic absolute value of $B_x$, and $L_z$ is the width of the box. We can
then calculate the other terms which do not involve trying to find the interface
and the parallel component of the electric field. The end result, which is the
new measure of reconnection rate, is
\begin{equation}
V_\mathrm{rec} = \frac{1}{2 |B_{x,\infty}| L_z} \left[ \oint{\mathrm{sign} (B_x) \vc{E} \cdot d \vc{l}} - \partial_t \int {|B_x| dA} \right]
\label{newmeasure}
\end{equation}

As one can note, this new reconnection measure contains the time derivative of
the absolute value of $B_x$, and a number of boundary terms, such as advection of
$B_x$ across the boundary and the boundary integral of the resistive term $\eta
\vc{j}$. The additional terms include all processes contributing the time change
of $|B_x|$. More discussion of the measure of reconnection given by Eq.~(\ref{newmeasure}), as well as results of numerical calculations of the individual terms entering the equation is given in Kowal, Lazarian \& Vishniac (2009).  

\section{Results}
\label{sec:results}

For the studies of the influence of turbulence on magnetic reconnection
obtained in the three dimensional simulations of the magnetic reconnection with the
measure given by Eq.~(\ref{oldmeasure}) we refer our reader 
to \cite{kowal08}. In this paper we compare the results obtained using the old and new measures,
i.e. compare results obtained with Eqs.~(\ref{oldmeasure}) and (\ref{newmeasure}). We show particular
examples of calculations obtained and defer a detailed parameter space study to Kowal et al. (2009) paper.

\begin{figure}[t]
\centering
\includegraphics[width=0.48\textwidth]{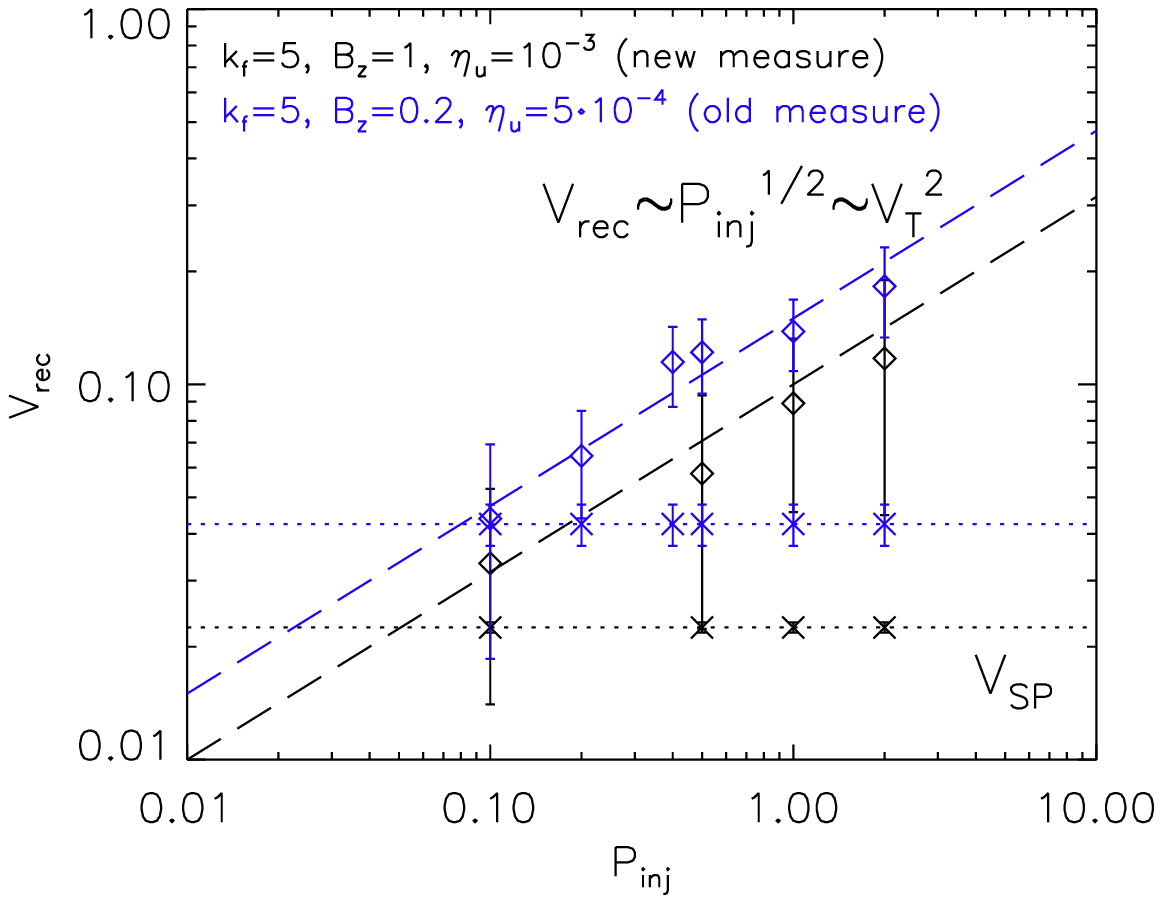}
\includegraphics[width=0.48\textwidth]{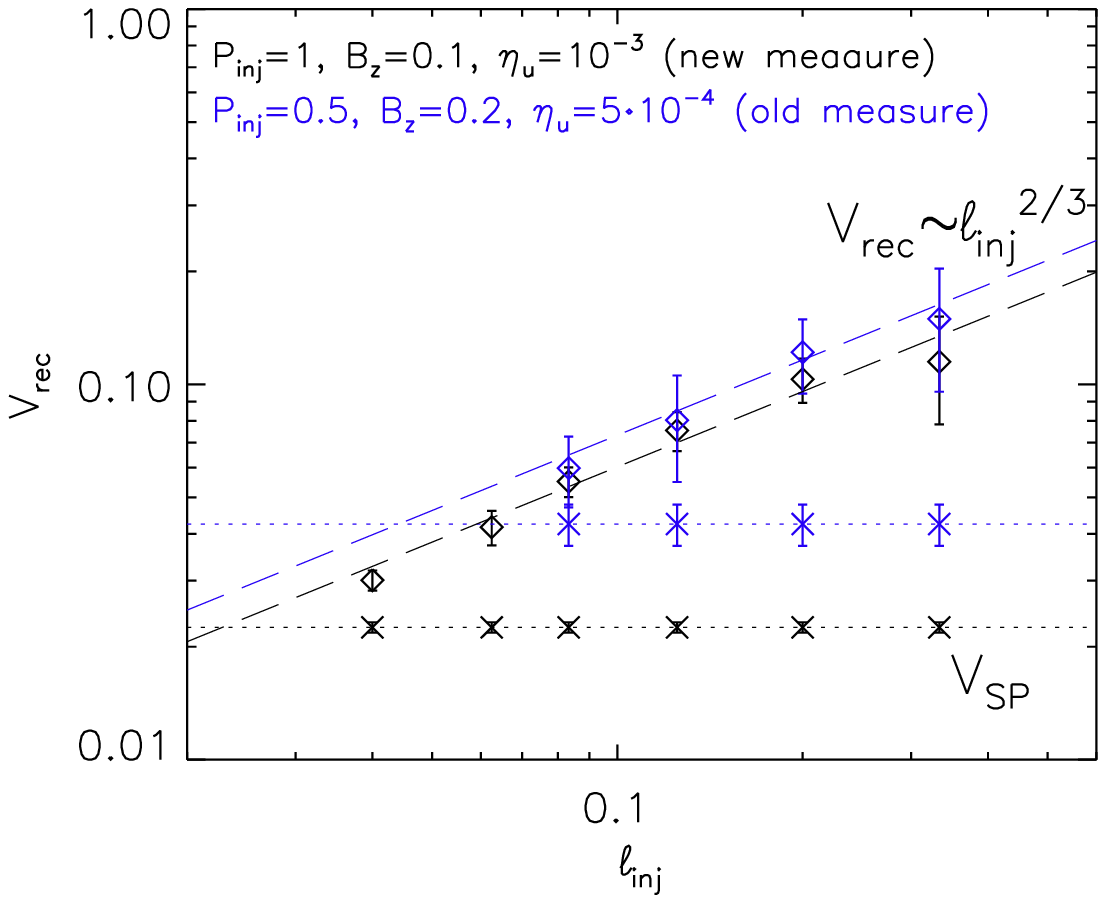}
\caption{The comparison of the old and new measure of the reconnection rate. {\it Left Panel} The dependence of the reconnection rate
$V_{rec}$ on the power of turbulence $P_{inj}$ (diamonds) using both approaches.
The corresponding Sweet-Parker rates, without the presence of turbulence, are
shown by x-symbols. {\it Right panel} The dependence of the
reconnection rate $V_{rec}$ on the injection scale $l_{inj}$ (diamonds). Again,
the Sweet-Parker rates are shown by x-symbols. \label{fig:deps}}
\end{figure}
In Figure~\ref{fig:deps} we show the comparison of the dependencies of the
reconnection rate obtained using old and new measure on the power of turbulence
(left plot) and the injection scale (right plot). In the case of turbulent power
dependency we used two different sets of models for each method. For the old
method of $V_{rec}$ estimation the shear component $B_z=0.2$, the uniform
resistivity $\eta=5 \cdot 10^{-4}$. The injection scale is the same for both
sets of models and $k_{inj}=5$. Dashed lines show the LV99 dependence $V_{rec}
\sim V_T^2$.

Both relations, using the old and new measures, show the same dependency, even
though they were fitted to two different sets of models. It means, that the
dependency of the reconnection rate is not sensitive to the strength of the
shear component $B_z$ as well as it stays unmodified with the change of uniform
resistivity and the numerical diffusivity of the method solving the MHD
equations\footnote{The observation that the old measure gives higher reconnection rate
can be explained by the fact that the set of models for the old measure were
calculated with a more dissipative code than the new set, which resulted in a 
higher value of numerical resistivity. Indeed, the Sweet-Parker
reconnection speed depends on the resistivity, the resulting values of $V_{rec}$
are higher.}

In the right plot of Figure~\ref{fig:deps} we show how  the old and
new measure of $V_{rec}$ depends on the injection scale
$l_{inj}$. Similarly to the dependence on the power of turbulence, this
dependence does not change and in both cases is $V_{rec} \sim l_{inj}^{2/3}$,
which is steeper than the $l^{1/2}$ rate predicted in LV99. However, LV99 assumed
a very simple model of turbulence no energy at scales large than $l$, while in our numerical 
studies we observe an inverse cascade and the energy is present at scales larger than $l$.
Therefore a steeper dependence that we observe is well justified. 
 
Thus, the most important conclusion that we get from our study is that the two measures of reconnection rate agree and provide results consistent with the LV99 predictions. In Kowal et al. (2008) we showed that the reconnection of the magnetic field in presence of turbulence is  independent of both anomalous and ordinary Ohmic resistivities.

\section{Discussion}
\label{sec:summary}

In this article we introduced a new way of measuring of reconnection rate and we
tested the dependence of the LV99 reconnection rate on the injection power and the injection scale
of MHD turbulence.  We found that the old and new reconnection measures produce consistent  
results and both confirm the predictions of the LV99 reconnection model.

We note that the numerical testing of the LV99 reconnection model is far from trivial. The
model is intrinsically three dimensional. In order to develop a turbulent cascade and minimize
the role of numerical diffusion we have
to use high resolution simulations. Another problem is the choice of proper
boundary conditions. Our model requires open boundaries in order to allow the
ejection of the reconnected flux.  This property is crucial for
the global reconnection constraint, since the reconnection stops when the
outflow of the reconnected flux is blocked.

If the turbulence level increases, LV99 model predicts that this accelerates magnetic reconnection. At very low levels of turbulence, the width over which magnetic fields wander gets smaller than the thickness of the Sweet-Parker current sheet and the Sweet-Parker reconnection takes over. 
 
The successful testing of LV99 model is good news for many areas of astrophysical numerical
modeling. Unlike the models that rely on particular plasma properties to induce fast reconnection,
the LV99 model is robust  and fast in any type of fluid, provided that 
the fluid is, at least, weakly turbulent. If the fluid is laminar initially, the LV99 model predicts that such magnetic configurations should be prone to bursts of reconnection, when the outflow increases the level of
turbulence in the system. The latter may provide an appealing explanation of Solar Flares (see more
in Lazarian \& Vishniac 2008). 

The LV99 model was used as a starting point for development of the model of reconnection of magnetic field in a partially ionized gas in Lazarian, Vishniac \& Cho (2004). Although the latter model has to be tested, our successes in testing of LV99 model is an encouraging sign in terms of understanding of reconnection in partially ionized interstellar gases. The latter process can serve provide removal of magnetic field during star formation, for instance (cf. Shu et al. 2007).

Other important implications of the fast reconnection of the weakly stochastic field include the acceleration of cosmic rays as magnetic fields lines shrink as a result of reconnection (see  de Gouveia dal Pino, E. \& Lazarian, A. 2003, 2005). This process is similar to acceleration of energetic particles considered in a more restrictive case of collisionless Hall-MHD reconnection by Drake et al. (2006).

\acknowledgements 
The research of GK and AL is supported by the Center for Magnetic Self-Organization in Laboratory and Astrophysical Plasmas and NSF Grant AST-0808118. The work of ETV is supported by the National Science and Engineering Research Council of Canada.  Part of this work was made possible by the facilities of the Shared Hierarchical Academic Research Computing Network (SHARCNET:www.sharcnet.ca). This research also was supported in part by the National Science Foundation through TeraGrid resources provided by Texas Advanced Computing Center (TACC:www.tacc.utexas.edu).


\end{document}